\newcommand{\bm}[1]{\mbox{{\it \boldmath$#1$}}}

\documentclass[cup5b]{caps}

\usepackage{graphicx}

\def\plotone#1{\centering \leavevmode
\includegraphics[width=.95\columnwidth]{#1}}

\begin{document}

\pagenumbering{arabic}

\author[]{MARIANGELA BERNARDI\\Carnegie Mellon University}

\chapter{Large Scale Structure in the Sloan Digital Sky Survey}

\begin{abstract}
The primary observational goals of the Sloan Digital Sky Survey are to 
obtain CCD imaging of 10,000 deg$^2$ of the north Galactic cap in five 
passbands, with a limiting magnitude in the $r$-band of 22.5, to obtain 
spectroscopic redshifts of $10^6$ galaxies and $10^5$ quasars, 
and to obtain similar data for three $\sim 200$ deg$^2$ stripes 
in the south Galactic cap, with repeated imaging to allow co-addition 
and variability studies in at least one of these stripes. 
The resulting photometric and spectroscopic galaxy datasets allow one 
to map the large scale structure traced by optical galaxies over a 
wide range of scales to unprecedented precision.  Results relevant to 
the large scale structure of our Universe include:  
a flat model with a cosmological constant $\Omega_\Lambda=0.7$ provides 
a good description of the data;  
the galaxy-galaxy correlation function shows departures from a power law 
which are statistically significant; 
and galaxy clustering is a strong function of galaxy type.  
\end{abstract}

\section{Introduction to the SDSS}
The Sloan Digital Sky Survey (SDSS; York et al. 2000) is the result 
of an international collaborative effort which includes scientists from 
the U.S., Japan and Germany (see {\tt http://www.sdss.org} for details).  
In brief, the survey uses a dedicated 2.5 meter telescope located at 
the Apache Point Observatory in New Mexico.  
Images are obtained by drift scanning with a mosaic camera of 
30 2048x2048 CCDs positioned in six columns and five rows (Gunn et al. 1998), 
which gives a field of view of $3\times 3~\mathrm{deg^2}$, 
with a spatial scale of $0.4~\mathrm{arcsec\,pix^{-1}}$ in five bandpasses 
($u$, $g$, $r$, $i$, $z$) with central wavelengths 
(3560, 4680, 6180, 7500, 8870\AA) (Fukugita et al. 1996).
The effective exposure time is 54.1 seconds through each CCD.
The SDSS image processing software provides several global photometric
parameters for each object, which are obtained independently in each 
of the five bands. The data are flux-calibrated by comparison with a set of 
overlapping standard-star fields calibrated with a 0.5-m ``Photometric 
Telescope''. 

The SDSS takes spectra only for a target subsample of calibrated imaging 
data (Strauss et al. 2002).
Spectra are obtained using a multi-object 
spectrograph which observes 640 objects at once.
The wavelength range of each spectrum is $3800-9200$~\AA. 
The instrumental dispersion is $\log_{10}\lambda=10^{-4}$dex/pixel 
which corresponds to 69~km~s$^{-1}$ per pixel.
Each spectroscopic plug plate, 1.5 degrees in radius, has 640 fibers, 
each $3~\mathrm{arcsec}$ in diameter. Two fibers cannot be closer than 
$55~\mathrm{arcsec}$ due to the physical size of the fiber plug.  
Typically $\sim 500$ fibers per plate are used for galaxies, 
$\sim 90$ for QSOs, and the remaining for sky spectra and 
spectrophotometric standard stars.  

At the time of writing, the SDSS had imaged roughly $\sim 4,500$ square 
degrees; $\sim 265,000$ galaxies and $\sim 35,000$ QSOs had both photometric 
and spectroscopic information. The first 460 square degrees and 50,000
spectra have been made public in an Early Data Release (see 
Stoughton et al. 2002, which includes many technical details of the 
survey), and roughly four times this will be made available in early 2003.

Data from the multi-waveband SDSS has already made significant 
contributions to our knowledge of the structure of our Milky Way 
galaxy and its satellites, 
correlations between galaxy observables, such as luminosity, size, 
velocity dispersion, color, chemical composition, star-formation rate, 
etc., and how these depend on galaxy environment, active galactic 
nuclei, high redshift quasars, the Ly$\alpha$ forest and the epoch 
of reionization.  
But in this article I will focus exclusively on published results from 
the SDSS about the large-scale structure of the Universe.  

\section{Galaxy clustering}
In the most successful theoretical models, galaxies grew by gravitational 
instability from initial seed fluctuations which left their imprint 
on the CMB.  The statistics of these initial fluctuations are expected 
to be Gaussian, so that complete information about these fluctuations 
is encoded in the shape of the power spectrum $P(k)$ of the initial 
density fluctuation field.  Nonlinear gravitational instability is 
expected to modify the shape of $P(k)$, and to make the fluctuation 
field at the present time rather non-Gaussian.  These changes are 
expected to be less severe on large scales, although, because gravity 
must compete with the expansion of the universe, what is meant by 
`large' depends on the amplitude of the initial fluctuations and 
on the background cosmology.  Thus, the large-scale distribution of 
galaxies at the present time encodes a wealth of cosmological information; 
one of the principal scientific goals of the SDSS collaboration is to 
extract this information.  
On smaller scales, the clustering is sensitive to the nonlinear 
gastrophysics of galaxy formation.  A generic prediction of most 
galaxy formation models is that clustering should be a strong 
function of galaxy type:  more luminous galaxies are expected to be 
more strongly clustered.  The SDSS database is ideally suited to 
quantifying how clustering depends on galaxy properties.  

With this in mind we discuss measures of clustering in the SDSS 
angular photometric catalogs first.  Although these lack the 
three-dimensional information present in redshift surveys, so most 
of the clustering signal is washed out by projection effects, angular 
catalogs are competitive because they have so many more galaxies 
than spectroscopic surveys.  
Section~\ref{ang} presents the angular two-point functions, 
$\omega(\theta)$ and $C_\ell$, measured in various apparent magnitude 
limited catalogs drawn from the SDSS database.  These can be thought 
of as measurements of the three dimensional power spectrum $P(k)$ 
through different windows.  
It then shows the result of inverting these measurements to derive 
constraints on the shape and amplitude of $P(k)$.  
Constraints on $P(k)$ which were obtained more directly from the 
angular data, without first estimating $w(\theta)$ or $C_\ell$, 
are also described.  

Section~\ref{3d} presents results from the three dimensional catalogs.  
These are considerably sparser, since spectra are only taken for 
objects with $r$-band magnitudes less than about 17.5, whereas the 
photometry is complete to $r<22.5$.  
Measurements of clustering in these are complicated by the fact that 
we only measure the redshift of a galaxy, not the comoving distance 
to it---the measured redshift depends both on the distance to the 
object and the component of its motion along the line-of-sight.  
Therefore, measures of clustering in redshift space are distorted 
compared to clustering in real space.  If motions are driven by 
gravity alone, then the difference depends on cosmology in a predictable 
way---at least on very large scales.  Although the data available at 
present do not probe these large scales, when the survey is complete, 
the SDSS dataset will provide an exquisite test of whether or not 
gravitational instability is the sole source of large scale motions.  
On the smaller scales ($<$ 15 Mpc) probed by the present data, 
galaxy clustering is a strong function of galaxy type---this is highlighted 
in Section~\ref{3d}.  Moreover, the SDSS measurements clearly show that 
the two-point correlation function of galaxies, long described as a 
simple power-law, does in fact show a statistically significant 
feature on scales of a few Mpc.  

One of the great virtues of the accurate multi-band photometry of the 
SDSS is that it allows one to make reasonably precise estimates of 
galaxy redshifts for most objects even when spectra are not available.  
Measurements of clustering in these photometric redshift catalogs 
provide the benefit of large galaxy numbers associated with the 
photometric catalogs, while the photometric redshift estimate can be 
used to reduce the amount by which the clustering signal is washed-out 
by projection.  Moreover, since the photometric catalog is considerably 
deeper than the spectroscopic one, it allows one to probe the evolution 
of clustering out to considerably higher redshifts.  
These measurements offer a promising way of estimating the evolution 
of clustering out to redshifts of order unity.  

For want of space, I only present results from the lowest order measures 
of clustering: two point statistics.  Higher-order clustering measures 
such as the moments of counts-in-cells (Szapudi et al. 2002) and the 
void distribution, the bispectrum, the $n$-point correlation functions, 
and topological measures such as the genus (Hoyle et al. 2002) and 
other Minkowski functionals have also been, or currently are being 
studied.  
The high quality of the SDSS data also allows various measurements 
of the weak gravitational lensing effect:  McKay et al. (2002) describe 
galaxy-galaxy shear measurements, and projects to study galaxy-galaxy 
and galaxy-quasar magnification bias are underway.  
Also, Nichol et al. (2000) and Bahcall et al. (2002) describe what has 
been learnt from galaxy clusters in the SDSS, and what the future holds
for such studies.  

\section{Angular clustering}\label{ang}
In theory, the two point correlation function $\omega(\theta)$ and 
the angular power spectrum $C_\ell$ are Fourier (actually Legendre) 
transforms of one another.  Therefore, in theory, they contain the same 
information.  In practice, incomplete sky coverage and other complications 
mean that the measured values of these two quantities are not equivalent, 
so the SDSS collaboration has measured both.  

\subsection{The angular correlation function $\omega(\theta)$}\label{wt}
In studies of large scale structure, galaxies are treated as points, 
and the statistics of point processes are used to quantify galaxy 
clustering.  One of the simplest of these statistics is the two-point 
correlation function which measures the excess number of (galaxy) 
pairs, relative to an unclustered (Poisson) distribution, as a function 
of pair separation.  Operationally, the two point correlation function 
is estimated by generating an unclustered random catalog with the same 
geometry as the survey, and then measuring 
$$
\omega(\theta) \equiv {DD - 2DR - RR\over RR}
$$
where $DD$, $DR$ and $RR$ are data--data, data--random, and 
random--random pair counts in bins of $\theta+\delta\theta$ 
in the data and random catalogs.  

\begin{figure}
 \plotone{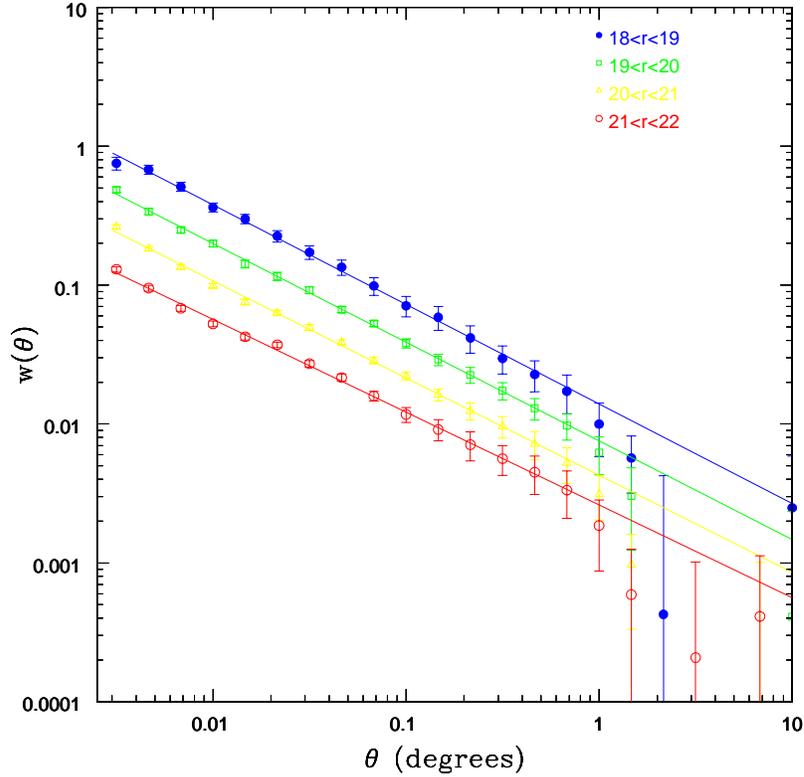}
 \caption{The angular correlation function $\omega(\theta)$ in 
  different magnitude limited catalogs drawn from the SDSS database
  (from Connolly et al. 2002).}
 \label{wtheta}
\end{figure}

Previous measurements of $\omega(\theta)$, primarily from wide-field 
photographic plate surveys of the sky, have shown that $\omega(\theta)$ 
is rather well-fit (at least at small separations) by a power law:  
$\omega(\theta)= (\theta/\theta_0)^{1-\gamma}$ with 
$\gamma\approx 1.7$.  The clustering amplitude, characterized by 
$\theta_0$, is expected to depend on the depth of a magnitude limited 
survey such as the SDSS.  
This is because the galaxy distribution is expected to be clustered 
isotropically in three-dimensions.  A photometric catalog projects out 
the radial component of the pair separation; the same angular 
separation can result from galaxies which have vastly different 
radial separations.  Since the clustering amplitude is smaller on 
large separations, a deeper catalog contains more pairs which are 
close in the direction perpendicular to the line-of-sight but are 
well-separated along the line-of-sight, thus diluting the overall 
clustering signal.

Figure~\ref{wtheta} shows $\omega(\theta)$ for SDSS galaxies in 
a number of different apparent magnitude bins.  The solid lines 
show power-law fits to the data, over the range $1'<\theta<30'$ 
(the fits use the full covariance matrix from Scranton et al. 2002).  
Notice that the angular clustering signal on large scales is small:  
at one degree, $\omega(\theta)\sim 0.013$ for galaxies with 
$18<r^*<19$.  Therefore, sky-position dependent errors in photometric 
calibration could dominate the signal.  Scranton et al. (2002) 
describes the results of a battery of tests designed to quantify, 
and where possible correct for,  the effects of photometric errors, 
stellar contamination, seeing, extinction, sky brightness, bright 
foreground objects and optical distortions in the camera itself.  
These tests highlight one of the great features of the SDSS 
dataset---its uniformity.  

Notice that a power law is a good but not perfect description of the 
data.  Also, the fainter catalogs, which contain galaxies out to greater 
distances, have a smaller angular clustering amplitude.  The precise 
scaling with apparent magnitude depends on cosmology:  
a flat universe with $\Lambda=0.7$ provides a much cleaner scaling 
than does one in which $\Omega_m=1$ (but we have not shown this here).  
As a rough guide to the scales involved, note that the median redshift 
of galaxies with $18<r^*<19$ is $z_m=0.18$ (this median redshift is 
0.24, 0.33 and 0.43 for the successively fainter galaxy catalogs).  
In a flat universe with $\Lambda=0.7$, one arcminute at $z=0.18$ 
corresponds to a distance of 154 $h^{-1}$kpc, so that 1 $h^{-1}$Mpc 
subtends about 0.11 degrees.  
Clearly, this estimate of $\omega(\theta)$ probes clustering on 
rather small scales.  The next section describes an estimate of the 
clustering strength on larger scales.  

\subsection{The angular power spectrum $C_\ell$}\label{cl}
There are three good reasons for computing the angular power spectrum 
$C_\ell$ in addition to the angular correlation function 
$\omega(\theta)$.  First, on large scales, where the Gaussian 
approximation is most likely to apply, the $C_\ell$ estimators retain 
all of the information contained in the angular clustering signal.  
Therefore, they represent a lossless compression of the full data set.  
Second, although both $\omega(\theta)$ and $C_\ell$ are obtained by 
averaging the three dimensional power spectrum $P(k)$ over window 
functions, say $W_\theta(k)$ and $W_\ell(k)$, the second of these, 
$W_\ell$, is considerably narrower.  This is advantageous if, as we 
will do shortly, one wishes to invert the measured two-dimensional 
statistic so as to constrain the form of $P(k)$.  
Narrow window functions are particularly important since small scale 
clustering is expected to be highly non-Gaussian; if the window 
function is broad, then one must worry about aliasing from small 
scale power.  
And third, it is possible to produce measurements of $C_\ell$ in 
which errors are uncorrelated.  

\begin{figure}
 \plotone{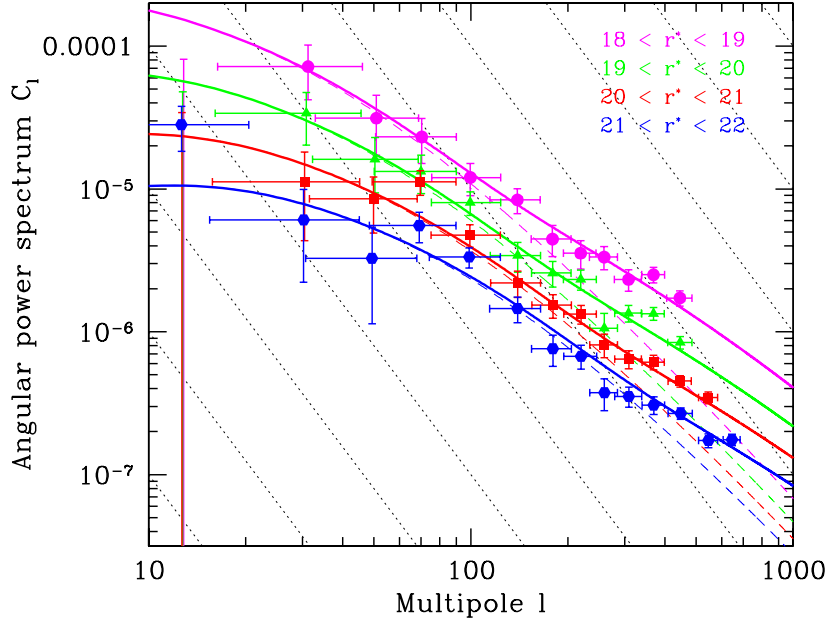}
 \caption{The angular power spectrum $C_\ell$ in different magnitude   
  limited catalogs drawn from the SDSS database.  As discussed in the 
  text, a crude estimate of the underlying three dimensional power 
  spectrum is obtained by shifting the same curve vertically and 
  horizontally by an amount which depends on the survey depth:  
  apparently fainter and more distant galaxies should be shifted 
  farther up (because there is more averaging along the line-of-sight 
  which has supressed fluctuations) and to the left (because as the 
  survey depth increases a given angular scale $\ell$ corresponds to 
  larger spatial scales).  Dotted lines show the direction of this 
  shift when the survey depth is changed (from Tegmark et al. 2002).}
 \label{csubls}
\end{figure}

Briefly, the measurement is made by dividing the sky patch into $N$ 
square pixels each $12'.5$ on a side and computing the density 
fluctuation $\delta_i=n_i/\bar n_i - 1$ in each pixel.  
These $\delta_i$s can be grouped into a vector $\bm{d}$, 
the covariance matrix of which is 
$$
{\bm C} \equiv \Bigl\langle {\bm d}{\bm d'}\Bigr\rangle = {\bm S} + {\bm N}
\qquad{\rm where}\qquad {\bm S} = \sum_i p_i {\bm P}_i.
$$
Here ${\bm N}$, assumed to be a diagonal matrix, denotes the 
contribution to ${\bm C}$ which comes from the fact that the galaxy 
distribution is discrete (this is sometimes called the shot-noise 
contribution), and $p_i$ denotes the parameters which specify the 
amplitude of the power spectrum, and the ${\bm P}_i$ are matrices which 
are specified by the survey geometry in terms of Legendre polynomials.  

The next step is to determine the $p_i$s from the observed data vector 
${\bm d}$.  This involves repeatedly multiplying and inverting 
$N\times N$ matrices, which is computationally expensive.  Therefore 
the Karhunen--Lo\`eve method is used to compress the information 
content of the map before estimating the power spectrum parameters.  
The actual estimates are made using a quadratic estimator which 
effectively Fourier transforms the sky map, squares the Fourier modes 
in the $i$th power spectrum band, and averages the results together.  
The details of this procedure are described in Tegmark et al. (2002).  

The results are shown in Figure~\ref{csubls}.  
A multipole $\ell$ corresponds roughly to an angular scale 
$\theta\sim 180^\circ/\ell$, so that $\ell=600$, for 
galaxies at $z=0.18$, corresponds roughly to a spatial scale of 
order 3 $h^{-1}$Mpc.  

\subsection{Inversion to the three-dimensional $P(k)$}\label{pk}
The previous sections presented estimates of the angular correlation 
function and power spectrum from the SDSS database.  
These measurements can be used to derive constraints on the three 
dimensional power spectrum.  This is possible because the angular power 
spectrum is related to the three dimensional power spectrum by 
$$
 C_\ell = \int_0^\infty {dk\over k}\,k^3 P(k)\, W_\ell(k),\qquad 
 {\rm where}\qquad 
 W_\ell(k) = {2\over\pi}\left[\int_0^\infty dr\,f(r)\,j_\ell(kr)\right]^2.
$$
Here $f$ is the probability distribution for the comoving distance $r$ 
to a random galaxy in the survey (which, in a photometric survey, is 
{\it not} measured), $j_\ell$ is a spherical Bessel function, and we 
have ignored the fact that the power spectrum evolves with redshift 
(strictly speaking, this expression also makes the standard assumption 
that clustering does not depend on luminosity;  
the next section shows that the data do not support this assumption, 
but the quantitative effect on the following analysis is small).  
To see what the definition above implies, note that for large values 
of $\ell$, corresponding to small angular scales, 
$j_\ell(kr)$ is sharply peaked around $kr=\ell$.  
Assuming the unknown $f(r)$ varies smoothly, we can set it equal to 
$f(\ell/k)$ and take it out of the integral above, leaving an integral 
over $j_\ell$ only which can be evaluated analytically.  
Thus, in this approximation, 
$\ell^3 W_\ell(k)\to [(\ell/k)\,f(\ell/k)]^2$, 
and 
$$C_\ell \to \int_0^\infty {dk\over k}\,{k^3 P(k)\over \ell^3}\,
            \left[{\ell\over k} \,f\!\left({\ell\over k}\right)\right]^2
         \approx {k_\ell^3 P(k_\ell)\over \ell^3}\int_0^\infty {dk\over k}\,
              \left[{\ell\over k} \,f\!\left({\ell\over k}\right)\right]^2,
$$
where the second approximation comes from assuming that the term in 
square brackets is sharply peaked about its mean value $k_\ell$.  
This term depends on the distribution of comoving distances.  
To see how, define $r_*\equiv\int dr\,rf(r)$.  
Then the assumption that $f$ is peaked means we should set 
$r_* \equiv\beta\ell/k_\ell$, where $\beta$ is a constant of order unity.  
Thus, $C_\ell \approx (k_\ell/\ell)^3 P(k_\ell) 
              \approx (\beta/r_*)^3 P(\beta\ell/r_*)$.
In other words, $C_\ell$ is a smoothed version of $P(k)$, 
which is shifted vertically (by a factor $r_*^3$) and horizontally 
(by $r_*$) on a log-log plot, by an amount which depends on the depth 
of the sample.  

On small scales, the angular correlation function is also related 
to the power spectrum by a window function:  
$$
\omega(\theta) = \int {dk\over k}\, k^2P(k)\,W_\theta(k)
\quad {\rm where}\quad 
W_\theta(k) = {1\over 2\pi}\int dr\, J_o(kr\theta)\,f^2(r),
$$
and $J_o$ is a Bessell function.
In contrast to the window associated with $C_\ell$, this $W_\theta$ 
oscillates around zero, so that it is harder to associate a single 
wave number $k$ with the angular $\omega(\theta)$.  Nevertheless, one 
can still develop techniques for inverting the measured $\omega(\theta)$ 
and $C_\ell$ to obtain the form of $P(k)$.  A number of these methods 
are summarized in Dodelson et al. (2002).  

\begin{figure}
 \plotone{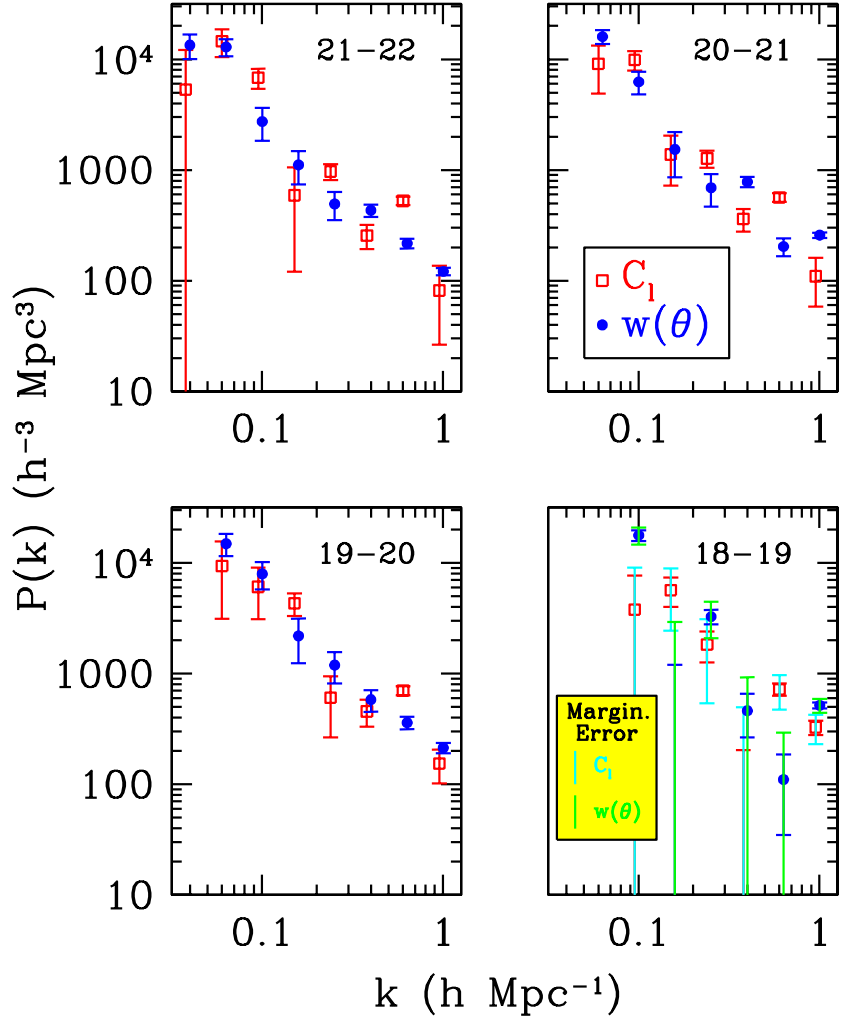}
 \caption{Comparison of $P(k)$ obtained by inverting $\omega(\theta)$ 
  and $C_\ell$ in four apparent magnitude bins.  Error bars are 
  unmarginalized; marginalizing over the non-zero covariances between 
  $k$ bins makes the error bars much larger (from Dodelson et al. 2002). }
 \label{pkinvert}
\end{figure}

Clearly, the results of the inversion are sensitive to the assumed form 
for $f(r)$ which, in turn, depends on the magnitude limit of the sample 
and cosmology.  (This is because, at fixed redshift, a flat model with 
$\Lambda>0$ has more volume than when $\Lambda=0$.  Therefore, if the 
sample depth is characterized by a median redshift, which is observable 
at least in principle, then the typical physical separation between 
galaxies at that redshift is larger in a flat model with $\Lambda>0$.)  
Figure~\ref{pkinvert} shows the result of inverting $\omega(\theta)$ 
and $C_\ell$ in the four apparent magnitude limited samples presented 
earlier; the inversion assumed a flat model with $\Omega_\Lambda=0.7$.  
Over the range of scales where they overlap, the estimates agree with 
one another.  

The implications for $P(k)$ are usually expressed as constraints on 
the parameters $\sigma_{8g}=b\sigma_8$ and $\Gamma$ which describe the 
amplitude (i.e. an up-down shift of all the points in Figure~\ref{pkinvert}) 
and the shape (how far to the right does $P(k)$ peak).  
The subscript $g$ indicates that the measured $P(k)$ is of the 
galaxy distribution rather than the dark matter, and the factor $b$ 
comes from the standard assumption (consistent with numerical 
simulations of clustering on large scales) that the power in the two 
distributions differs only by a multiplicative linear bias factor.  
As the figure shows, the strongest constraints on the shape parameter 
come from the faintest galaxies (i.e. the magnitude bin $21 < r < 22$):
$\Gamma = 0.14^{+0.11}_{-0.06}$ ($95\%$ C.L.).  
The shape of $P(k)$ also depends on the baryon fraction 
$\Omega_b/\Omega_m$: increasing this ratio suppresses power on scales 
smaller than the peak, and analysis of the full data set will set 
interesting limits on this parameter also.  

\subsection{Direct estimates of $P(k)$}\label{pkkl} 
The previous subsection described estimates of the three dimensional 
$P(k)$ which were derived by first measuring projected quantities 
$\omega(\theta)$ and $C_\ell$.  Since these are essentially smoothed 
versions of $P(k)$, an alternative procedure is to circumvent the 
initial measurement of projected quantities, and to work instead with 
quantities which optimize the signal-to-noise of the dataset.  
This is the KL approach taken by Szalay et al. (2002) who first 
expand the projected galaxy distribution on the sky over a set of 
Karhunen-Lo\`eve eigenfunctions, and then use a maximum likelihood 
analysis to derive constraints on the shape and amplitude of $P(k)$.  
For a flat universe with a cosmological constant, they find 
$\Gamma=0.188\pm 0.04$ and $\sigma_{8g} = 0.915 \pm 0.06$ 
(statistical errors only).  Since $\Gamma\approx\Omega_mh$, 
if we use the HST measurement of the Hubble constant to set $h=0.7$, 
then the SDSS results imply $\Omega_m\approx 0.27$.


\section{Clustering in $z$ space}\label{3d}
The spectroscopic sample provides galaxy redshifts, and hence a 
reasonably accurate distance measurement, so that, in contrast to 
the angular photometric catalogs, a much stronger clustering signal 
can be measured.  Moreover, because the redshift is available, it is 
possible to derive an accurate estimate of the intrinsic luminosity 
of each galaxy.  This allows one to estimate how clustering depends 
on intrinsic, rather than apparent, properties of galaxies such as 
luminosity and rest-frame color.  This is important because, in 
magnitude limited surveys like the SDSS, the most luminous galaxies 
are visible at the greatest distances, whereas the least luminous 
galaxies are only visible nearby.  
Therefore, the power on the largest scales is dominated by the 
clustering of the most luminous galaxies, whereas the power on smaller 
scales comes from a mix of galaxy types.  If clustering depends on 
luminosity, then one must account for the changing mix of galaxy types 
at each scale when estimating the shape of the power spectrum.

\begin{figure}
 \plotone{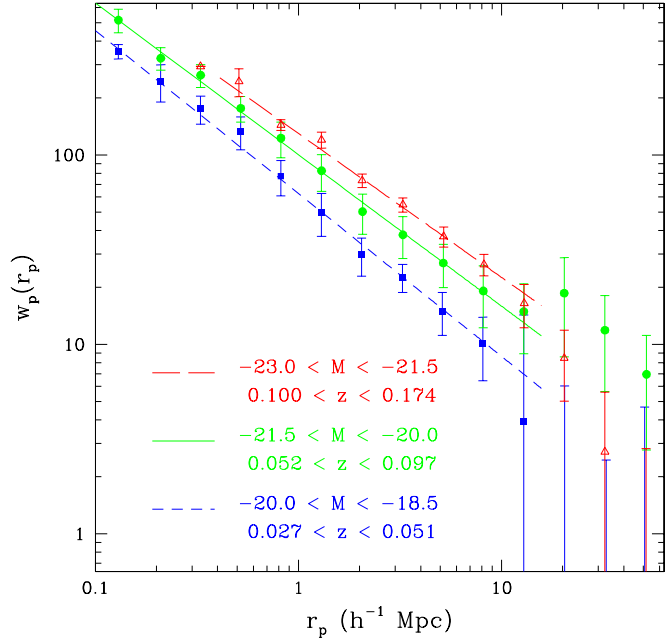}
 \caption{Galaxy clustering depends on luminosity.  
  Changing the luminosity changes the amplitude but not the slope of the 
  correlation function (from Zehavi et al. 2002).}
 \label{xil}
\end{figure}

As mentioned previously, peculiar velocities distort clustering 
statistics in redshift space.  One way of accounting for these 
distortions is to measure the correlation function as a function of 
the separation parallel and perpendicular to the line-of-sight:  
$\xi_2(r_p,\pi)$.  
Only separations parallel to the line-of-sight $\pi$ are affected by 
peculiar motions, so that 
$$
w_p(r_p) \equiv 2\int_0^\infty d\pi\,\xi_2(r_p,\pi)
              = 2\int_0^\infty d\pi\,\xi\Bigl(\sqrt{r_p^2 + \pi^2}\Bigr)
$$
is independent of redshift-space distortions.  Since 
$$
P(k_p,k_\pi) \equiv \int dr_p \int d\pi\, \xi_2(r_p,\pi) \,
               \exp(-i k_p r_p - i k_\pi\pi), 
$$
the quantity $P(k_p,0)$, being the Fourier transform of 
$w_p(r_p)$, is also distortion free.  

\begin{figure}
 \plotone{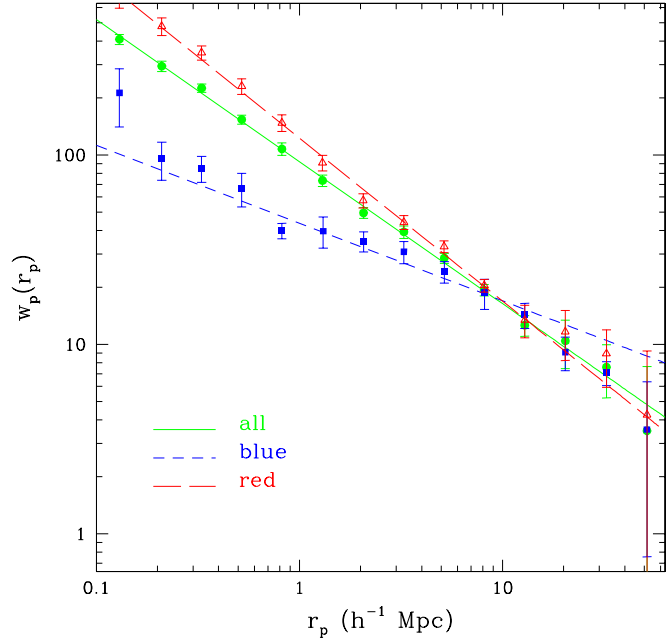}
 \caption{Galaxy clustering depends on color.  
  Changing the color changes the slope and amplitude of $\xi(r)$ 
  (from Zehavi et al. 2002).}
 \label{xic}
\end{figure}

Measurements of the distortion-free correlation function and power 
spectrum both show that more luminous galaxies are more strongly 
clustered than less luminous galaxies (Figure~\ref{xil}).  
Whereas the amplitude of $\xi(r)$ appears to depend strongly on 
luminosity, the shape is approximately independent of $L$.
On the other hand, the shape of the correlation function depends 
strongly on color:  redder galaxies have steeper correlation functions 
(Figure~\ref{xic}).

As discussed by Budavari et al. (2003), these trends are qualitatively 
consistent with the following simple model.  Suppose there are two 
types of galaxies, each with their own clustering pattern (say, a red 
population with a steeper correlation function than the blue population).  
Then the correlation function of the entire sample will be a weighted 
sum of the two populations, the weighting being determined by the 
relative numbers of the two types.  
Next, suppose that the amplitude of the correlation function in 
subsamples of each population depends on luminosity, and that the scaling 
with luminosity is similar for the two populations (on scales larger 
than 1~Mpc, this is a good description of the SDSS data; on smaller 
scales, clustering strength increases with luminosity for blue galaxies, 
whereas red galaxies show the opposite trend).  
Finally, suppose that the luminosity functions of these two populations 
have similar shapes, at least at the luminous end.  These requirements 
guarantee that the correlation functions of subsamples defined by 
luminosity will always have the same shape, whereas subsamples defined 
differently will have different shapes.  
Explaining why this should be so is an interesting challenge for 
galaxy formation models.  

A first study of clustering using photometric redshifts provides 
qualitatively similar results (Budavari et al. 2003).  
This is extremely encouraging because the use of photometric redshifts 
allows one to go considerably fainter than the spectroscopic dataset 
allows.  In particular, photo-$z$s offer a cost-effective way of probing 
clustering out to redshifts of order unity.  That this is possible 
at all is a tribute to the accuracy of the SDSS photometry.  

Since the full $\xi(r_p,\pi)$ is sensitive to peculiar velocities, 
whereas $w_p(r_p)$ is not, a comparison of the two provides a 
measurement of galaxy peculiar velocities.  On the small scales to 
which the present data is most sensitive, the dependence of clustering 
on luminosity and type constrains the velocity dispersions of the halos 
which different galaxy types populate.  The SDSS data show that 
early-type galaxies populate halos with larger velocity dispersions 
(Zehavi et al. 2002), in qualitative agreement with the fact that 
such galaxies are much more common in massive clusters than in the 
field.  

Since the first measurements of Totsuji \& Kihara (1969), the galaxy 
correlation function has been characterized as a power law.  
A look through most of the figures presented here shows that, while a 
power law is indeed a good description, it is not perfect.  The SDSS 
correlation functions show rich structure, much of which is statistically 
significant.  In most ab initio models of $\xi(r)$, power-laws are purely 
fortuitous---they are not generic.  Explaining the positions of the bumps 
and wiggles and their dependence on galaxy type, and hence extracting 
information from these features in $\xi(r)$ will become a rich area 
of research. 

Funding for the creation and distribution of the SDSS Archive has been provided by the Alfred P. Sloan Foundation, the Participating Institutions, the National Aeronautics and Space Administration, the National Science Foundation, the U.S. Department of Energy, the Japanese Monbukagakusho, and the Max Planck Society. The SDSS Web site is http://www.sdss.org/.


\begin{thereferences}{}

\bibitem{} Bahcall, N. A., Feng, D., Bode, P., et al. 2002, ApJ, in press (astro-ph/0205490)
 
\bibitem{} Budavari, T., Connolly, A. J., Szalay, A. S., et al. 2003 (in preparation)

\bibitem{} Connolly, A. J., Scranton, R., Johnston, D., et al. 2002, ApJ, 579, 42

\bibitem{} Dodelson, S., Narayanan, V. K., Tegmark, M., et al. 2002, ApJ, 572, 140

\bibitem{} Fukugita, M., Ichikawa, T., Gunn, J. E., et al. 1996, AJ, 111, 1748

\bibitem{} Gunn, J.E., Carr, M.A., Rockosi, C.M., Sekiguchi, M., et al. 1998, AJ, 116, 3040

\bibitem{} Hoyle, F.,  Vogeley, M. S., Gott III, J. R. 2002, ApJ, 580, 663

\bibitem{} McKay, T. A., Sheldon, E. S., Racusin, J., et al. 2002, ApJ, submitted (astro-ph/0108013)

\bibitem{} Nichol, R., Miller, C., Connolly, A., et al. 2000, HEAD meeting 32, 14.02 (astro-ph/0011557)
 
\bibitem{} Scranton, R., Johnston, D., Dodelson, S., et al. 2002, ApJ, 579, 48

\bibitem{} Stoughton, C., Lupton, R.H., Bernardi, M., et al. 2002, AJ, 123, 485 (Early Data Release)

\bibitem{} Strauss, M.A., Weinberg, D.H., Lupton, R.H. et al. 2002, AJ, 124, 1810

\bibitem{} Szalay, A. S., Jain, B., Matsubara, T., et al. 2002, ApJ, in press (astro-ph/0107419)

\bibitem{} Szapudi, I., Frieman, J. A., Scoccimarro, R. 2002, ApJ, 570, 75

\bibitem{} Tegmark, M., Dodelson, S., Eisenstein, D. J., et al. 2002, ApJ, 571, 191

\bibitem{} Totsuji, H. \& Kihara, T. 1969, PASJ, 21, 221

\bibitem{} York, D.G., Adelman, J., Anderson, J.E., et al. 2000, AJ, 120, 1579

\bibitem{} Zehavi, I., Blanton, M., Frieman, J., et al. 2002, ApJ, 571, 172

\end{thereferences}
\end{document}